%% file: IdEhMimo_AkAs.tex
\documentclass[10pt,journal]{IEEEtran}

\usepackage{amsmath}
\usepackage{amssymb}
\usepackage{nicefrac}
\usepackage{graphicx}
\usepackage{pgfplots}
\usepackage{algorithm}
\usepackage{algpseudocode}
\usepackage{cite}
\usepackage{subfigure}
\usepackage{balance} 
\usepackage{stfloats}
\usetikzlibrary{shapes,snakes}
\usetikzlibrary{calc}
\usetikzlibrary{patterns}
\usetikzlibrary{decorations.pathmorphing} 

\usepackage[normalem]{ulem}

\newcommand{\RxPowerSplitting}[4]{
	\coordinate (a) at (#1,#2);
	\draw[line width=0.25pt,scale=(#3)] (a)--($(a)+(-0.2,0)$)--($(a)+(-0.2,0.7)$)--
	($(a)+(-0.1,0.8)$)--($(a)+(-0.3,0.8)$)--($(a)+(-0.2,0.7)$);
	\draw ($(a)+(0.15,0)$) circle (0.15cm);
	\node at ($(a)+(0.15,0)$) {\small +};
	\draw ($(a)+(0.3,0)$)--($(a)+(0.5,0)$);
	\draw[->] ($(a)+(0.15,-0.5)$)--($(a)+(0.15,-0.15)$);
	\node at ($(a)+(0.15,-0.7)$) {$w_{#4}$};
	\draw[fill=yellow!20] ($(a)+(0,-0.3)+(0.5,0)$) rectangle ($(a)+(1,0.2)+(0.5,0.1)$);
	\node at ($(a)+(0.5,0)+(0.5,0)$){\footnotesize $\text{PS}_{#4}$};
}
\newcommand{\RxAntennaNoise}[4]{
	\coordinate (a) at (#1,#2);
	\draw[line width=0.25pt,scale=(#3)] (a)--($(a)+(-0.2,0)$)--($(a)+(-0.2,0.7)$)--
	($(a)+(-0.1,0.8)$)--($(a)+(-0.3,0.8)$)--($(a)+(-0.2,0.7)$);
	\draw ($(a)+(0.15,0)$) circle (0.15cm);
	\node at ($(a)+(0.15,0)$) {\small +};
	\draw ($(a)+(0.3,0)$)--($(a)+(0.5,0)$);
	\draw[->] ($(a)+(0.15,-0.5)$)--($(a)+(0.15,-0.15)$);
	\node at ($(a)+(0.15,-0.7)$) {$w_{#4}$};
}

\newcommand{\TxAntenna}[3]{
	\coordinate (a) at (#1,#2);
	\draw[line width=0.25pt,scale=(#3)] (a)--($(a)+(0.2,0)$)--($(a)+(0.2,0.7)$)--
	($(a)+(0.1,0.8)$)--($(a)+(0.3,0.8)$)--($(a)+(0.2,0.7)$);
}

\newcommand{\RxAntenna}[4]{
	\coordinate (a) at (#1,#2);
	\draw[line width=0.25pt,scale=(#3)] (a)--($(a)+(-0.2,0)$)--($(a)+(-0.2,0.7)$)--
	($(a)+(-0.1,0.8)$)--($(a)+(-0.3,0.8)$)--($(a)+(-0.2,0.7)$);
}
\begin{document}

	\title{Towards Optimal Energy Harvesting Receiver Design in MIMO Systems}
	\author{
		\IEEEauthorblockN{Ali Kariminezhad, \textit{Student Member, IEEE}, and Aydin Sezgin, \textit{Senior Member, IEEE}}\\
		
	}
	\maketitle
\thispagestyle{empty}
\pagestyle{empty}	
	
\begin{abstract}
In this paper, we investigate a multiple-input multiple-output (MIMO) system with simultaneous information detection (ID) and energy harvesting (EH) receiver. This point-to-point system operates in the vicinity of active interfering nodes. The receiver performs power splitting where a portion of received signal undergoes analog energy harvesting circuitry. Further, the information content of the other portion is extracted after performing digital beamforming. In this MIMO system, information carrier eigen-modes are not necessarily the eigen-modes with the strongest energy level. Hence, it is beneficial to perform independent beamforming at the receiver of MIMO-P2P channel. Here, we utilize a hybrid analog/digital beamforming for the purpose of simultaneous ID and EH in such scenarios. This design, provides extra design degrees-of-freedom in eigen-mode selection for ID and EH purposes independently. Worst-case performance of this receiver structure is discussed. Finally, its benefits are compared to the classical receiver structure and the gains are highlighted.  
\end{abstract}
	\vspace*{-0.3cm}
\section{introduction}
Wireless power transfer is regarded as one of the new features future communication systems are expected to offer~\cite{Baszynski2016}. This has to be provided simultaneous to the ever increasing demand of information transfer~\cite{Krikidis2014}.  Hence, optimal transceiver design for simultaneous wireless information and power transfer (SWIPT) is of crucial importance. In a single-antenna receiver two approaches are common by now. One approach is that the information of the received radio frequency (RF) signal is extracted in a time instant, however the energy is harvested in another time instant. Alternatively, by power splitting a portion of received signal is conveyed to the ID chain and the other portion passes to the energy harvesting circuit~\cite{Shi2014}. In a simple manner, a rectifier converts the energy of the incident RF signal to direct current (DC), which in turn loads the energy buffer. This energy can thus help the user stay in the network actively longer~\cite{Hameed2014}.

Now, in order to enhance the information transmission rate, multiple antennas can be deployed at the transmitter and receiver. This multiple-input multiple-output (MIMO) channel has significantly higher capacity compared to the single-input single-output channel~\cite{Telatar1999}. Considering SWIPT, the authors in~\cite{Zhang2013} study MIMO broadcast systems, where the users either demand information or energy. The authors in~\cite{Liu2013} investigate multi-antenna power splitting receivers for maximizing the achievable rate. In that paper, the authors apply power splitting after constructive combining of the received signals from the antennas in the analog domain. While in~\cite{Kariminezhad2017SPL} a hybrid beamforming structure for simultaneous ID and EH in a single-input multiple-output P2P channel was proposed. In this paper, we study the performance of simultaneous ID and EH in a MIMO communication system. We consider a point-to-point (P2P) MIMO in the proximity of another multiple antenna interference source. This interference source can be for instance, a high power base station (BS) which serves the users in a cell, or a WiFi node for in-home or short-range applications. Two potential PS structures are depicted in~Fig.~\ref{fig:ReceiverStructure}. As shown in~Fig.~\ref{fig:a1} the power splitter first splits the signal from each antenna separately which then proceeds with the beamforming process. Therefore, the P2P receiver has the degrees of freedom to perform independent digital and analog beamforming for information detection and energy harvesting purposes, respectively. However, in~Fig.~\ref{fig:b1}, the receiver first combines the received signal from multiple antennas, then the received signal undergoes power splitter for simultaneous ID and EH. Since energy of an analog signal can be captured, the signal combining is inevitably performed in analog domain. Here, we consider a single analog chain for energy harvesting purposes, however multiple digital processing chains are considered for information reception. Considering the PS structure shown in~Fig.~\ref{fig:a1}, we analyze the system performance from the achievable rate and harvested energy perspective for the two following network architectures, 1) classical information transfer and 2) simultaneous wireless information and power transmission (SWIPT).
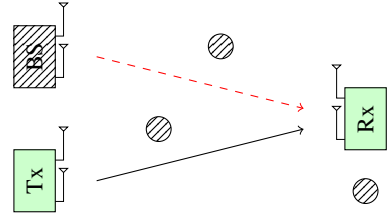
\begin{figure}
			\centering
			\begin{tikzpicture}[scale=0.55, every node/.style={scale=0.9}]
			
			\draw[fill=yellow!20,pattern=north east lines] (-8,-1)rectangle(-7,0.5);
			\node[rotate=90] at (-7.5,-0.25){BS};
			\TxAntenna{-7}{0.25}{1}
			\TxAntenna{-7}{-0.75}{1}
			
			\draw[fill=green!20] (-8,-4)rectangle(-7,-2.5);
			\node[rotate=90] at (-7.5,-3.25){Tx};
			\TxAntenna{-7}{-3.75}{1}
			\TxAntenna{-7}{-2.75}{1}
			
			\draw[fill=green!20] (0,-2.5)rectangle(1,-1);
			\node[rotate=90] at (0.5,-1.75){Rx};
			\RxAntenna{0}{-1.25}{1};
			\RxAntenna{0}{-2.25}{1};
			
			\draw[->] (-6,-3.25)--(-1,-2);
			\draw[->,dashed,red] (-6,-0.25)--(-1,-1.5);
			
			\draw[fill=yellow!20,pattern=north east lines] (-3,0) circle (0.3cm);
			\draw[fill=yellow!20,pattern=north east lines] (-4.5,-2) circle (0.3cm);
			\draw[fill=yellow!20,pattern=north east lines] (0.5,-3.5) circle (0.3cm);
			\end{tikzpicture}
			\caption{P2P-MIMO in the proximity of a multi-antenna transmitter. This transmitter can be a BS serving cellular users, for instance.}
			\label{fig:SystemModel}
		\end{figure}
\section{System Model}
	Consider the deployment of a multiple-input multiple-output point-to-point (MIMO-P2P) pair of nodes in a cellular downlink with a multiple-antenna base station (BS) as depicted in~\figurename{~\ref{fig:SystemModel}}. The received signal at the P2P receiver equipped with $K$ antennas is given by
	\begin{align}
	\mathbf{y}=\mathbf{H}\mathbf{x}+\mathbf{H}_\text{B}\mathbf{x}_\text{B}
	+\mathbf{w},
	\end{align}
	where $\mathbf{y}\in\mathbb{C}^{K}$ is the complex-valued received signal vector. The channel matrix from the transmitter of the P2P pair and from the base station are represented by $\mathbf{H}\in\mathbb{C}^{K\times M}$ and $\mathbf{H}_\text{B}\in\mathbb{C}^{K\times N}$, respectively. Notice that, the number of transmit antennas at the transmitter of the P2P channel is denoted by $M$, and BS is equipped with $N$ antennas. The transmit signal from the P2P transmitter and from the BS to the P2P receiver are denoted by $\mathbf{x}\in\mathbb{C}^{M}$ and $\mathbf{x}_\text{B}\in\mathbb{C}^{N}$, respectively. The receiver additive white Gaussian antenna noise (AWGN) vector is denoted by $\mathbf{w}$. Here, we assume that, $K\leq \min(M,N)$. This assumption limits the degrees-of-freedom (DoF) by the number antennas at the receiver. Here, DoF is defined as the number of interference-free parallel data streams between the transmitter and the receiver.\\
	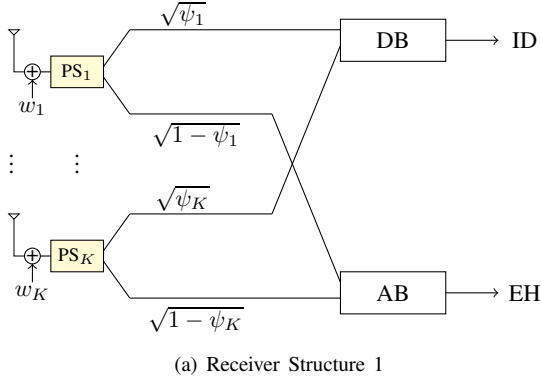
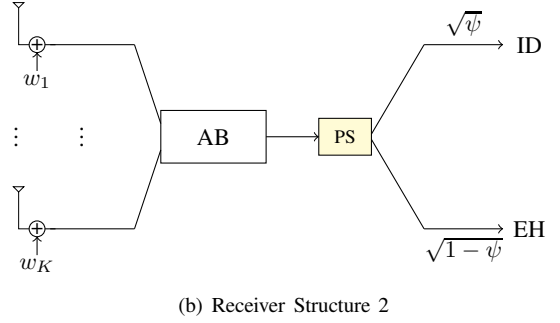
\begin{figure*}
			\centering
			\begin{minipage}[b]{0.45\textwidth}
					\centering
					\subfigure[\footnotesize Receiver Structure 1]{
						\tikzset{every picture/.style={scale=0.7}, every node/.style={scale=0.9}}%
						\input{PsThenBF}
						\label{fig:a1}
			}
			\end{minipage}\quad\quad\quad\quad
			\begin{minipage}[b]{0.45\textwidth}
			\centering
			\subfigure[\footnotesize Receiver Structure 2]{
				\tikzset{every picture/.style={scale=0.7}, every node/.style={scale=0.9}}%
				\input{BfThenPS}
				\label{fig:b1}
			}
			\end{minipage}
			\caption{Simultaneous information detection (ID) and energy harvesting (EH) at a multi-antenna receiver by power splitting (PS). Digital and analog beamformings are depicted by DB and AB, respectively.}
			\label{fig:ReceiverStructure}
		\end{figure*}
	The receiver is capable of simultaneous information detection (ID) and energy harvesting (EH) by power splitting (PS). Here, we adopt the hybrid beamforming structure of~Fig.~\ref{fig:a1} which allows, a) the freedom of utilizing MIMO multiplexing gain which is beneficial from achievable rate perspective, b) independent analog beamforming which is beneficial for energy harvesting.
	The information symbols at the P2P user and BS are linearly precoded to obtain the transmit signal as
	\begin{align}
	\mathbf{x}_\text{B}&=\sum_{n=1}^{N}\mathbf{v}_{\text{B}_n}d_{\text{B}_n}=
	\mathbf{V}_{\text{B}}\mathbf{d}_{\text{B}},\\
	\mathbf{x}&=\sum_{m=1}^{M}\mathbf{v}_m d_m=\mathbf{V}\mathbf{d},
	\end{align}
	respectively. Notice that $\mathbf{v}_{\text{B}_n}$ is the beamforming vector dedicated to the $n$th single-antenna cellular user. Furthermore, $M$ corresponds to the number of transmit chains at P2P transmitter. At the energy harvesting-capable (EHC) receiver, power splitting is performed first. Let $\sqrt{\psi_k}$ portion of the signal captured by the $k$th received antenna undergoes ID chain and $\sqrt{1-\psi_k}$ portion passes to the EH circuit. Then, the signal after digital beamforming is formulated by
	\begin{align}
	\mathbf{z}_{\text{ID}}=\mathbf{U}^{H}\boldsymbol{\Psi}\mathbf{y}+
	\mathbf{U}^{H}\mathbf{n},
	\end{align}
	where $\mathbf{U}$ is the beamforming matrix at the P2P receiver and $\boldsymbol\Psi=\text{diag}\left[\sqrt{\psi_1}, ..., \sqrt{\psi_K} \right]$. Notice that, $\mathbf{n}$ is the aggregate noise of the ID process which is modeled as AWGN, i.e., $\mathbf{n}\sim\mathcal{CN}(\mathbf{0},\sigma^{2}_n\mathbf{I})$. The signal after analog beamforming is given by
	$
	\mathbf{z}_{\text{EH}}=\mathbf{q}^{H}\boldsymbol{\Theta}\mathbf{y},
	$
	where $\mathbf{q}$ is the analog beamfoming vector. Moreover, $\boldsymbol{\Theta}=\text{diag}\left[\sqrt{1-\psi_1}, ..., \sqrt{1-\psi_K} \right]$. Now, the channel input-output relationship is given by
	\begin{align}
	\mathbf{z}_\text{ID}=&\mathbf{U}^{H}\boldsymbol{\Psi}\left(\mathbf{H}
	\mathbf{V}\mathbf{d}+\mathbf{H}_\text{B}
	\mathbf{V}_\text{B}\mathbf{d}_\text{B}+\mathbf{w}\right)+\mathbf{U}^{H}\mathbf{n},\label{zId}\\
	\mathbf{z}_{\text{EH}}=&\mathbf{q}^{H}\boldsymbol{\Theta}\left(\mathbf{H}
	\mathbf{V}\mathbf{d}+\mathbf{H}_\text{B}
	\mathbf{V}_\text{B}\mathbf{d}_\text{B}+\mathbf{w}\right).\label{zEh}
	\end{align}
	In what follows, we analyze the two cases, namely information only transmission, and SWIPT in details.

	\section{Classical Information Transfer}
	In this section, we analyze the performance of the joint ID and EH receiver, where the transmitters employ classical information transfer.

	\subsection{Information Detection}
	We define equivalent channels $\hat{\mathbf{H}}=\boldsymbol{\Psi}\mathbf{H}$ and $\hat{\mathbf{H}}_\text{B}=\boldsymbol{\Psi}\mathbf{H}_\text{B}$. Furthermore, we perform singular value decomposition (SVD), $\hat{\mathbf{H}}=\hat{\mathbf{L}}\hat{\boldsymbol\Lambda}\hat{\mathbf{R}}^{H}$. Then, we can reformulate equation~\eqref{zId} as
	\begin{align}
	\mathbf{z}_\text{ID}=&\mathbf{U}^{H}\left(\hat{\mathbf{L}}
	\hat{\boldsymbol\Lambda}\hat{\mathbf{R}}^{H}\mathbf{V}\mathbf{d}+
	\hat{\mathbf{H}}_\text{B}\mathbf{V}_\text{B}\mathbf{d}_\text{B}+\hat{\mathbf{w}}
	+\mathbf{n}\right),\label{zId2}
	\end{align}
	where $\hat{\mathbf{w}}=\boldsymbol\Psi \mathbf{w}$.
	In this section, we consider the availability of global channel state information, and local channel state information at the MIMO-P2P.
	For the general analysis, we proceed with the assumption that the global CSI is available at the transmitter and receiver sides of the P2P pair. Worst-case analysis is presented when only local CSI is available at the P2P pair. This analysis is only studied for comparison purpose.
	\subsubsection{Global CSI}
	Having the global channel knowledge and by treating the incident interference as noise (TIN) at the receiver, the achievable rate is given by~\cite{Telatar1999}
	\begin{align}
	r_\text{G-csi}\left(\mathbf{Q}\right)=\log\det\left(\mathbf{I}
	+\hat{\mathbf{H}}^{H}\mathbf{S}^{-1}\hat{\mathbf{H}}\mathbf{Q}\right),
	\end{align}
	where $\mathbf{S}=\hat{\mathbf{H}}_\text{B}\mathbf{Q}_\text{B}
	\hat{\mathbf{H}}^{H}_\text{B}+\boldsymbol\Sigma_{\hat{w}}+\boldsymbol\Sigma_n$. The transmit covariance matrices of the P2P transmitter and BS are represented by $\mathbf{Q}=\mathbf{V}^{H}\mathbb{E}\{\mathbf{d}\mathbf{d}^{H}\}\mathbf{V}$ and $\mathbf{Q}_\text{B}=\mathbf{V}^{H}_\text{B}\mathbb{E}\{\mathbf{d}_\text{B}
	\mathbf{d}_\text{B}^{H}\}\mathbf{V}_\text{B}$, respectively. Moreover, the equivalent antenna and processing noise covariance matrices are denoted by $\boldsymbol{\Sigma}_{\hat{\mathbf{w}}}=\sigma^{2}_{w}
	\boldsymbol{\Psi}^{2}$ and $\boldsymbol{\Sigma}_{\mathbf{n}}=\sigma^{2}_{n}\mathbf{I}$, respectively. Notice that, for any given $\mathbf{Q}_\text{B}$ and $\boldsymbol{\Psi}$, the achievable rate is a function of $\mathbf{Q}$, which is maximized by $
	\mathbf{Q}=\mathbf{G}\left(\eta\mathbf{I}-\mathbf{M}^{-1} \right)^{+}\mathbf{G}^{H},$
	where $\hat{\mathbf{H}}^{H}\mathbf{S}^{-1}\hat{\mathbf{H}}=
	\mathbf{G}\mathbf{M}\mathbf{G}^{H}$,~\cite{Ye2003}. Furthermore, the optimal power allocation is obtained by water-filling, so that the water lever $\eta$ is adjusted to satisfy the power constraint, i.e., $\text{Tr}\left(\eta\mathbf{I}-\mathbf{M}^{-1} \right)=P$.
	\subsubsection{Local CSI}
	Here, we study the optimal transceiver design with the local CSI available at the P2P pair for ID purpose. This is realized by transmit and receive beamforming in the direction of the left and right singular-value matrices of $\hat{\mathbf{H}}$. That means, $\mathbf{U}=\hat{\mathbf{L}}$ and $\mathbf{V}=\hat{\mathbf{R}}$. Therefore, we obtain,
	\begin{align}
	\mathbf{z}_\text{ID}=&\hat{\boldsymbol\Lambda}\mathbf{d}+
	\mathbf{U}^{H}\left(\hat{\mathbf{H}}_\text{B}
	\mathbf{V}_\text{B}\mathbf{d}_\text{B}+\hat{\mathbf{w}}\right)+
	\mathbf{U}^{H}\mathbf{n}.\label{zId3}
	\end{align}
	Now, given $\mathbf{U}$, $\mathbf{V}$ and considering that the BS resources (power and antenna dimensions) are not devoted to help P2P pair, we obtain
	$
	\mathbf{z}_\text{ID}=\hat{\boldsymbol\Lambda}\mathbf{d}+
	\boldsymbol{\Gamma}\mathbf{d}_\text{B}+\bar{\mathbf{w}}+
	\bar{\mathbf{n}},
	$
	where $\boldsymbol{\Gamma}=\mathbf{U}^{H}\hat{\mathbf{H}}_\text{B}
	\mathbf{V}_\text{B}$, $\bar{\mathbf{w}}=\mathbf{U}^{H}\hat{\mathbf{w}}$ and $\bar{\mathbf{n}}=\mathbf{U}^{H}\mathbf{n}$.
	Now the achievable information rate over the P2P channel with the imposed interference from the BS can be formulated as
	\begin{align}
	r_\text{L-csi}=\log\det\left(\mathbf{I}+\bar{\mathbf{S}}^{-1}\hat{\boldsymbol{\Lambda}}
	\mathbf{D}\hat{\boldsymbol{\Lambda}}^{H}\right),\label{AchiRate}
	\end{align}
	where $\bar{\mathbf{S}}=\boldsymbol{\Gamma}\mathbf{D}_\text{B}\boldsymbol{\Gamma}^{H}
	+\boldsymbol{\Sigma}_{\bar{w}}+\boldsymbol{\Sigma}_{\bar{n}}$. Notice that, the incident interference is treated as noise (TIN). Moreover, the matrices $\mathbf{D}=\mathbb{E}\{\mathbf{d}\mathbf{d}^{H}\}$ and $\mathbf{D}_\text{B}=\mathbb{E}\{\mathbf{d}_\text{B}\mathbf{d}^{H}_\text{B}\}$ correspond to the covariance matrices of the information symbols at the P2P transmitter and BS, respectively.\\
	 \begin{figure*}[hb!]
	 	       \hrule
	 	   	   \begin{align}
	 	   	   p^\star_{\text{B}_k}=\frac{1}{\hat{\lambda}^{2}_{\text{B}_k}}\left(\sqrt{\left(\beta_k+\frac{\alpha_k}{2}\right)^2+\left(\frac{\alpha_k\hat{\lambda}^{2}_{\text{B}_k}}{\mu}-\beta^2_k-\alpha_k\beta_k\right)}-\left(\beta_k+\frac{\alpha_k}{2}\right)^2 \right),\quad\forall k\tag{18}\label{solutionA}
	 	   	   \end{align}
	\end{figure*}
	\textit{Worst-case interference}:
	Let $\hat{\mathbf{L}}_{\text{B}}
	\hat{\boldsymbol\Lambda}_{\text{B}}\hat{\mathbf{R}}^{H}_{\text{B}}$ be the SVD of $\hat{\mathbf{H}}_{\text{B}}$. Then, the left singular-value matrix of the worst-case channel matrix from the BS to the P2P receiver, i.e., $\hat{\mathbf{L}}_{\text{B}}$, and the worst-case transmit beamforming matrix at the BS, i.e., $\hat{\mathbf{V}}_{\text{B}}$, have the following properties~\cite{Jorswieck2004},
	\begin{align}
\hat{\mathbf{L}}_{\text{B}}=\mathbf{U}=\hat{\mathbf{L}},\quad
	\mathbf{V}_{\text{B}}=\hat{\mathbf{R}}_{\text{B}}.
	\end{align}
	Then, the worst-case achievable rate can be written as
	\begin{align}
	r_{\text{wc}}&=\log\det\left( \mathbf{I}+ \left({\hat{\boldsymbol{\Lambda}}_\text{B}}
	\mathbf{D}_\text{B}
	{\hat{\boldsymbol{\Lambda}}^{H}_\text{B}} +\boldsymbol{\Sigma}_{\bar{w}}+\boldsymbol{\Sigma}_{\bar{n}} \right)^{-1} \hat{\boldsymbol{\Lambda}}
	\mathbf{D}\hat{\boldsymbol{\Lambda}}^{H} \right)\nonumber\\
	&=\sum_{k=1}^{K}\log\left(1+\frac{\hat{\lambda}^{2}_kp_k}{\hat{\lambda}^{2}_{\text{B}_k}p_{\text{B}_k} + \psi_k\sigma^{2}_{w_k} + \sigma^{2}_{n_k}} \right),
	\end{align}
	where $\hat{\lambda}_k$, $\hat{\lambda}_{\text{B}_k}$, $p_k$ and $p_{\text{B}_k}$ are the $k$th diagonal elements of $\hat{\boldsymbol{\Lambda}}$, $\hat{\boldsymbol{\Lambda}}_\text{B}$, $\mathbf{D}$ and $\mathbf{D}_\text{B}$, respectively.
	The achievable rate in~\eqref{AchiRate} is a function of the power allocation at the BS, i.e. $p_{\text{B}_k}$, and P2P transmitter, i.e. $p_k,\ \forall k$. Therefore, the power allocation problem needs to be considered for the worst-case achievable rate. We formulate the problem as
	\begin{subequations}\label{P1}
		\begin{align}
		\max_\mathbf{p}\ \min_{\mathbf{p}_{\text{B}}}\quad  r_\text{wc}\tag{\ref{P1}}\quad
		\text{s.t.}\quad &\sum_{k=1}^{K}{p_k}\leq P,\quad
		\sum_{k=1}^{K}{p_{\text{B}_k}}\leq P_\text{B},
		\end{align}
	\end{subequations}
	where $P$ and $P_\text{B}$ are the transmit power budget at the P2P transmitter and BS, respectively. Moreover, $\mathbf{p}=[p_1,...,p_K]$ and $\mathbf{p}_{\text{B}}=[p_{\text{B}_1},...,p_{\text{B}_K}]$. Notice that, the optimal power allocation at the P2P transmitter should correspond to the worst power allocation at the BS. Interestingly, the function $r_\text{wc}$ is concave in the optimization parameter vector $\mathbf{p}$ and convex in $\mathbf{p}_{\text{B}}$, moreover the constraint set is convex. This way the max-min optimization problem ~\eqref{P1} can be reformulated as a min-max optimization problem given by~\cite{Fan1953}
	\begin{subequations}\label{P1B}
			\begin{align}
			\min_{\mathbf{p}_{\text{B}}}\ \max_\mathbf{p}\quad   r_\text{wc}\tag{\ref{P1B}}\quad
			\text{s.t.}\quad &\sum_{k=1}^{K}{p_k}\leq P,\quad
			\sum_{k=1}^{K}{p_{\text{B}_k}}\leq P_\text{B},
			\end{align}
	\end{subequations}
	
	This satisfies the saddle-point property as well. Evidently, given the worst-case power allocation a the base station $\mathbf{p}^\star_\text{B}$ the optimal power allocation at the P2P transmitter is achieved by waterfilling with
	\begin{align}
	p^\star_k=\left(\eta-\frac{{\hat{{\lambda}}^{2}_{\text{B}_\text{k}}}
		{p}^\star_{\text{B}_\text{k}}
		+\psi_k\sigma^{2}_{w_k} + \sigma^{2}_{n_k}}{\hat{\lambda}^{2}_k}\right)^{+},\ \forall k,\label{waterfillA}
	\end{align}
	where the water level $\eta$ needs to be adjusted to fulfill the power constraint $P$. Now, given the optimal power allocation at the P2P transmitter, the function $r_\text{wc}$ is convex in $\mathbf{p}_\text{B}$. Hence, the dual problem provides the optimal solution of the primal problem. Assuming the optimal $\mathbf{p}^{\star}$ from~\eqref{waterfillA}, the Lagrangian of the minimization problem w.r.t. $\mathbf{p}_\text{B}$ is given by 
	\begin{align}
	\mathcal{L}(\mathbf{p}_\text{B},\mu)&=\sum_{k=1}^{K}\log\left(1+\frac{\alpha_k}{\hat{\lambda}^{2}_{\text{B}_k}p_{\text{B}_k} + \beta_k} \right)+\nonumber\\
	&\mu(\sum_{k=1}^{K}p_{\text{B}_k}-P_\text{B})+\sum_{k=1}^{K}\tau_k p_{\text{B}_k},
	\end{align}
    in which $\alpha_k=\hat{\lambda}^{2}_kp^\star_k$ and $\beta_k=\psi_k\sigma^{2}_{w_k} + \sigma^{2}_{n_k}$. Since, the optimization problem is convex w.r.t. $\mathbf{p}_\text{B}$, the necessary and sufficient
    Karush-Kuhn-Tucker (KKT) conditions hold. This results in 
    \begin{align}
    	\frac{\partial \mathcal{L}(\mathbf{p}_\text{B},\mu)}{p_{\text{B}_k}}&=\frac{-\alpha_k\hat{\lambda}^{2}_{\text{B}_k}}{\left(\hat{\lambda}^{2}_{\text{B}_k}p_{\text{B}_k} + \beta_k\right)\left(\hat{\lambda}^{2}_{\text{B}_k}p_{\text{B}_k} + \beta_k+\alpha_k\right)}+\mu=0.
    \end{align}
    From this equation, the worst-case power allocation at the base station is given at the bottom of the page in~\eqref{solutionA}, in which the Lagrangian multiplier $\mu$ is set, such that $\sum_{k=1}^{K}p^\star_{\text{B}_k}=P_\text{B}$.
\stepcounter{equation}
\begin{figure*}[t]
\centering
\begin{minipage}[b]{0.45\textwidth}
\subfigure[\footnotesize Receiver Structure 1, Worst-case vs. Average]{
\tikzset{every picture/.style={scale=1}, every node/.style={scale=0.9}}
\input{Rate1}
\label{fig:Rate1}
	}
\end{minipage}\quad\quad
\begin{minipage}[b]{0.45\textwidth}
\subfigure[\footnotesize Receiver Structure 1 vs. Receiver Structure 2]{
\tikzset{every picture/.style={scale=1}, every node/.style={scale=0.9}}
\input{Rate2}
\label{fig:Rate2}
	}
\end{minipage}\quad\quad\quad\quad\quad
            
\caption{Achievable rate comparison at the MIMO-P2P with the energy harvesting-capable receiver assuming classical information transfer.}
\end{figure*}
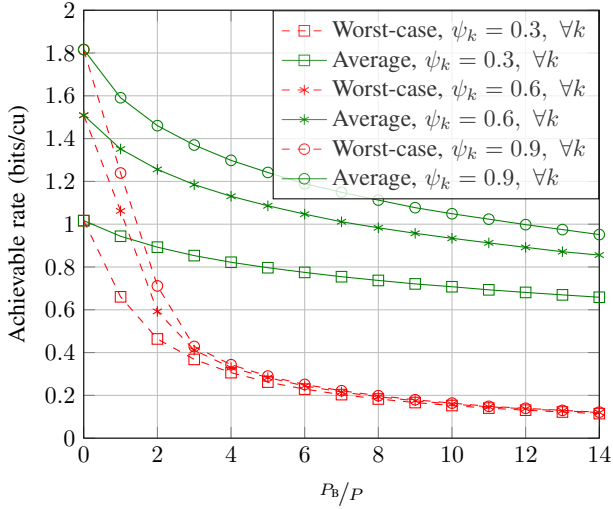
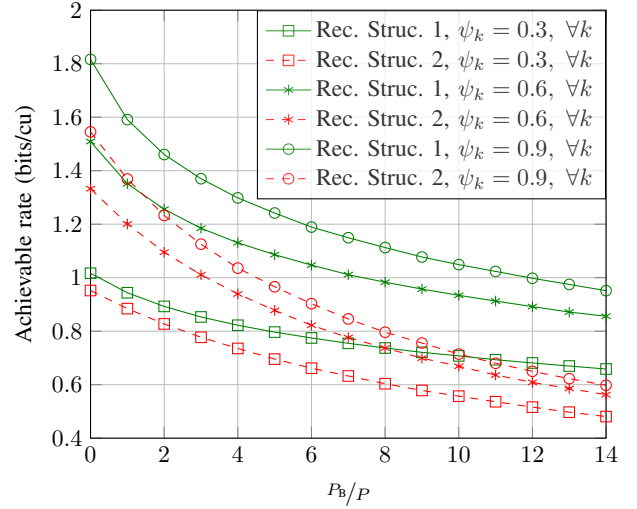
\subsection{RF Energy Harvesting}
As can be seen from~\eqref{zEh}, the amount of harvested energy is strongly dependent on the direction of analog steering vector $\mathbf{q}$. Hence, in order to maximize the harvested RF energy from the incident received signal, the steering vector $\mathbf{q}$ needs to be optimized. We define the received RF signal covariance matrix after power splitting by $\mathbf{C}_\text{RF}=\mathbf{C}+\mathbf{C}_\text{B}+\tilde{\mathbf{W}}$, where
\begin{align}
\mathbf{C}&=\mathbf{\Theta}\mathbf{H}\mathbf{V}\mathbf{D}\mathbf{V}^{H}\mathbf{H}^{H}\mathbf{\Theta}^{T},\\
\mathbf{C}_\text{B}&=\mathbf{\Theta}\mathbf{H}_\text{B}\mathbf{V}_\text{B}\mathbf{D}_\text{B}\mathbf{V}^{H}_\text{B}\mathbf{H}^{H}_\text{B}\mathbf{\Theta}^{T},\\
\tilde{\mathbf{W}}&=\sigma^{2}_w\boldsymbol{\Theta}^{2}.
\end{align} 
Notice that, $\mathbf{V}=\hat{\mathbf{R}}$, and $\mathbf{D}=\mathbb{E}\{\mathbf{d}\mathbf{d}^{H}\}$. Now, let the eigenvalue decomposition of $\mathbf{C}$ and $\mathbf{C}_\text{B}$ be, $\mathbf{C}=\tilde{\mathbf{L}}\tilde{\boldsymbol\Lambda}\tilde{\mathbf{L}}^{H}$ and $\mathbf{C}_\text{B}=\tilde{\mathbf{L}}_{\text{B}}
\tilde{\boldsymbol\Lambda}_{\text{B}}\tilde{\mathbf{L}}_{\text{B}}^{H}$. Then, the post-processed RF signal power is given by
\begin{align}
\mathbb{E}\{\mathbf{z}^{H}_{\text{EH}}\mathbf{z}_{\text{EH}}\}=&\mathbf{q}^{H}\left(
\tilde{\mathbf{L}}\tilde{\boldsymbol\Lambda}\tilde{\mathbf{L}}^{H}
+\tilde{\mathbf{L}}_{\text{B}}
\tilde{\boldsymbol\Lambda}_{\text{B}}\tilde{\mathbf{L}}_{\text{B}}^{H}
+\tilde{\mathbf{W}}\right)\mathbf{q},\label{zEh2}
\end{align}
where the optimal steering vector $\mathbf{q}$ corresponds to the direction of the maximum eigenvalue of all channels.
By defining $\tilde{\boldsymbol{\Lambda}}_{\text{B}_{i}}$ and $\tilde{\boldsymbol{\Lambda}}_{j}$ as the $i$th and $j$th diagonal elements of $\tilde{\boldsymbol{\Lambda}}_\text{B}$ and $\tilde{\boldsymbol{\Lambda}}$, respectively, we assume that
	\begin{align}
\tilde{\boldsymbol{\Lambda}}_{\text{B}_{1}}&\geq\tilde{\boldsymbol{\Lambda}}_{\text{B}_{i}}\ \forall i\in\big\{2,...,\min(M,N)\big\}\\
\tilde{\boldsymbol{\Lambda}}_{\text{B}_{1}}&\geq\tilde{\boldsymbol{\Lambda}}_{j},\forall j\in\big\{1,...,N\big\}.
	\end{align}
   Intuitively, this happens when a cellular user in the vicinity of P2P receiver obtains a strong signal from the BS. This way, $\mathbf{q}^{H}\tilde{\mathbf{L}}_B=[1,0,...,0]$. Therefore, the amount of the harvested energy is
	\begin{align}
	\mathbb{E}\{\mathbf{z}^{H}_{\text{EH}}\mathbf{z}_{\text{EH}}\}=
	\tilde{\lambda}_{\text{B}_1}+\mathbf{q}^{H}\left(\mathbf{C}+\sigma^{2}_w \boldsymbol{\Theta}^{2}\right)\mathbf{q},
	\end{align}
	where $\tilde{\lambda}_{\text{B}_1}$ is the first diagonal element of $\tilde{\boldsymbol\Lambda}_{\text{B}}$.
	In the next section, we consider the case, where the transmitters help the receiver not only with the information demands, but also with the energy demands.

\section{Simultaneous Information and Power Transfer (SWIPT)}
Here, we assume that the receiver has perfect knowledge of the energy signal. Thus, by interference cancellation, the energy signal can be completely removed at the information detection chain~\cite{Krikidis2014}. Hence, the achievable information rate does not deteriorate by the imposed interference from the energy signal. The transmit signal from the P2P transmitter and BS are given by $\mathbf{x}=\mathbf{V}^{\text{In}}\mathbf{d}^{\text{In}}+
\mathbf{V}^{\text{En}}\mathbf{d}^{\text{En}}$ and $
\mathbf{x}_\text{B}=\mathbf{V}^{\text{In}}_\text{B}\mathbf{d}^{\text{In}}_\text{B}+
\mathbf{V}^{\text{En}}_\text{B}\mathbf{d}^{\text{En}}_\text{B}$.
Now, having $\mathbf{d}^{\text{En}}$ and $\mathbf{d}^{\text{En}}_\text{B}$ at the P2P receiver, the achievable information rate is given by
\begin{align}
r\left(\mathbf{Q},\mathbf{Q}_\text{B}\right)=I(\mathbf{z}_\text{ID};\mathbf{x}|\mathbf{d}^{\text{En}},\mathbf{d}^{\text{En}}_\text{B})=\log\left(\mathbf{I}
+\hat{\mathbf{H}}^{H}\tilde{\mathbf{S}}^{-1}\hat{\mathbf{H}}\mathbf{Q}\right).\nonumber
\end{align}
			
Notice that, in order to maximize the achievable information rate, all signal power at the P2P transmitter needs to be allocated to the information signal. Furthermore, having the energy signals available at the P2P receiver, the energy signal transmission by the BS does not effect the achievable rate, hence all transmit power at the BS needs to be allocated to the energy signal. After interference cancellation, the overall noise covariance matrix is $\tilde{\mathbf{S}}=\boldsymbol\Sigma_{\hat{w}}+\boldsymbol\Sigma_n$. Assuming $\text{diag}(\tilde{\boldsymbol{\Lambda}}_\text{B})\succ_w\text{diag}(\tilde{\boldsymbol{\Lambda}})$, rate-energy optimal transmit signals are essentially
$\mathbf{x}=\mathbf{V}^{\text{In}}\mathbf{d}^{\text{In}}$ and $\mathbf{x}_\text{B}=\mathbf{V}^{\text{En}}_\text{B}\mathbf{d}^{\text{En}}_\text{B}$.

The optimality of this transmission can be intuitively justified. Maximum information rate can be achieved by allocating all transmit power and zero transmit power to the information signal at the P2P transmitter and BS, respectively. However, the harvested energy can be maximized with the maximum power allocation to the energy signal at the BS. This way, the optimal transmit covariance matrix at the P2P transmitter is $\mathbf{Q}=\mathbf{E}\left(\beta\mathbf{I}-\mathbf{N}^{-1} \right)^{+}\mathbf{E}^{H}$
where the eigenvalue decomposition of $\hat{\mathbf{H}}^{H}\tilde{\mathbf{S}}^{-1}\hat{\mathbf{H}}=
\mathbf{E}\mathbf{N}\mathbf{E}^{H}$. Furthermore, $\beta$ is the water level, which fulfills the power constraint at the P2P transmitter. The transmit covariance matrix of the optimal energy transmission from the BS is $
\mathbf{Q}_\text{B}=P_\text{B}\mathbf{e}_\text{B}\mathbf{e}^{H}_\text{B}$,
where $\mathbf{e}_\text{B}=\text{max-eig}\left(\boldsymbol{\Theta}\mathbf{H}_\text{B}\mathbf{H}^{H}_\text{B}
\boldsymbol{\Theta}^{T} \right)$. Notice that, the operator $\text{max-eig}$ denotes the eigenvector corresponding to the maximum eigenvalue of the input argument.

\begin{figure*}
\centering
\begin{minipage}[b]{0.45\textwidth}
\subfigure[\footnotesize Classical information transfer]{
\tikzset{every picture/.style={scale=1}, every node/.style={scale=0.9}}
\input{Energy1}
\label{fig:Energy1}
}
\end{minipage}\quad\quad
\begin{minipage}[b]{0.45\textwidth}
\subfigure[\footnotesize SWIPT]{
\tikzset{every picture/.style={scale=1}, every node/.style={scale=0.9}}
\input{Energy2}
\label{fig:Energy2}
		}
\end{minipage}\quad\quad
	            
\caption{Comparison between the harvested energy of the two receiver structures with both classical information transfer and SWIPT.}
\end{figure*}
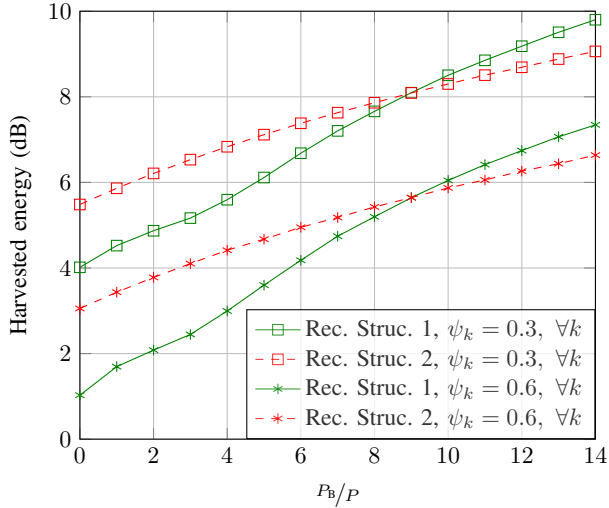
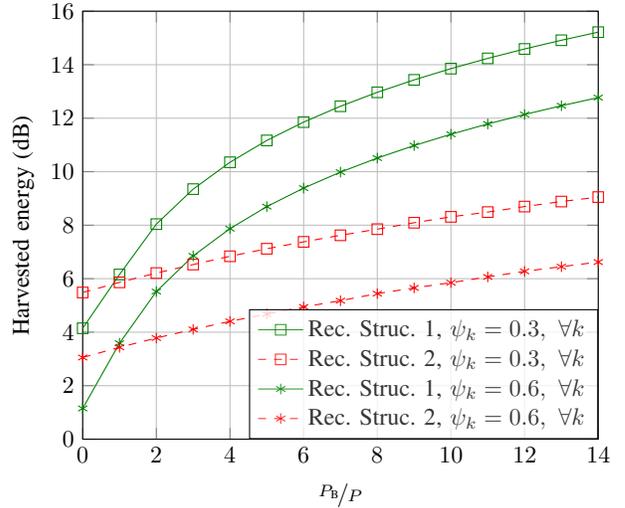
\section{Numerical Results}
In this section, we present the numerical results for the discussed network structures: 1) classical information transfer, 2) Simultaneous wireless information and power transfer (SWIPT). In our simulations, we assumed that the antenna noise and digital processing noise variances are equal to 1, i.e., $\sigma^{2}_w=\sigma^{2}_n=1$. The BS is equipped with 5 antennas, $N=5$. Moreover, the P2P transmitter and receiver each has 3 antennas, $K=3$. The non-zero singular values of the channel from the BS to the P2P transmitter are $[0.8,0.7,0.5]$. Furthermore, the eigenvalues of the channel from the P2P transmitter to its receiver are $[0.9,0.8,0.7]$. Fig.~\ref{fig:Rate1} depicts the achievable information rate as a function of transmit power ratio at the BS and P2P transmitter. We assumed that the BS serves the cellular users, which are distributed in the cell randomly. Hence, by random beamforming at the BS, we obtain the average rate of the P2P communication. As can be seen from~Fig.~\ref{fig:Rate1}, by increasing power ratio ($\nicefrac{P_\text{B}}{P}$), the gap between the worst-case and average achievable rates for $\psi_k=0.3,\ \forall k$, $\psi_k=0.6,\ \forall k$ and $\psi_k=0.9,\ \forall k$ decreases. This is due to the fact that, by $\nicefrac{P_\text{B}}{P}\gg 1$, the system operates in an interference-limited regime. In this regime, the distortion caused by the processing noise $n$, is negligible compared to the distortion from the interference signal. Moreover, the comparison between receiver structure presented in~Fig.~\ref{fig:ReceiverStructure} is shown in~Fig.~\ref{fig:Rate2}. Notice that, the transmitters perform classical information transfer and simultaneous ID and EH is performed at the multi-antenna receiver. We observe that, by utilizing receiver structure in~Fig.~\ref{fig:a1}, the achievable rate in the MIMO-P2P channel can be significantly enhanced. This is due to its capability in providing multi-streaming. The comparison in the harvested energy utilizing the two studied receiver structures is illustrated in~Fig.~\ref{fig:Energy1}. We observe that, at a particular interference regime, the classical receiver structure performs better from the harvested energy perspective. In the receiver structure in~Fig.~\ref{fig:b1}, due to single analog beamforming capability at the P2P receiver, the optimal information transmission is in the eigen-direction corresponding with the dominant eigen-value of the MIMO-P2P channel. Hence, the transmit energy is not distributed in multiple eigen-directions. Therefore, by single analog beamforming chain at the P2P receiver, higher amount of energy can be harvested. However, at high interference regime, the energy carried by interference eigen-direction becomes dominant and receiver structure~Fig.~\ref{fig:a1} performs better. As shown in~Fig.~\ref{fig:Energy2}, by SWIPT at the transmitters, the harvested energy of the receiver structure in~Fig.~\ref{fig:a1} performs significantly better that the structure in~Fig.~\ref{fig:b1}, which is due to its capability in independent beamforming for ID and EH purposes.    

\bibliographystyle{IEEEtran}
\bibliography{reference}
\balance
	
\end{document}

%% file: PsThenBF.tex
\begin{tikzpicture}

\RxPowerSplitting{-2}{0}{1}{1}
\node at (-2.25,-1.6){.};
\node at (-2.25,-1.75){.};
\node at (-2.25,-1.9){.};

\node at (-1,-1.6){.};
\node at (-1,-1.75){.};
\node at (-1,-1.9){.};
\RxPowerSplitting{-2}{-3.5}{1}{K}

\draw (-0.5,0.1)--(0,0.8);
\draw (0,0.8)--(4,0.8);
\node at (1,1.1){$\sqrt{\psi_1}$};

\draw (-0.5,-0.1)--(0,-0.8);
\draw (0,-0.8)--(2.7,-0.8);
\node at (1.25,-1.2){$\sqrt{1-\psi_1}$};
\draw (2.7,-0.8)--(4,-4);

\draw (-0.5,-3.4)--(0,-2.7);
\draw (0,-2.7)--(2.7,-2.7);
\node at (1,-2.4){$\sqrt{\psi_K}$};
\draw (2.7,-2.7)--(4,0.5);
\draw (4,0.2) rectangle (6,1);
\node at (5,0.6){DB};
\draw[->] (6,0.6)--(7,0.6);
\node at (7.5,0.6){ID};

\draw (-0.5,-3.6)--(0,-4.3);
\draw (0,-4.3)--(4,-4.3);
\node at (1.25,-4.7){$\sqrt{1-\psi_K}$};
\draw (4,-4.6) rectangle (6,-3.8);
\node at (5,-4.2){AB};
\draw[->] (6,-4.2)--(7,-4.2);
\node at (7.5,-4.2){EH};

\end{tikzpicture}

%% file: BfThenPS.tex
\begin{tikzpicture}

\RxAntennaNoise{-2}{0}{1}{1}
\node at (-2.25,-1.6){.};
\node at (-2.25,-1.75){.};
\node at (-2.25,-1.9){.};

\node at (-1,-1.6){.};
\node at (-1,-1.75){.};
\node at (-1,-1.9){.};
\RxAntennaNoise{-2}{-3.5}{1}{K}

\draw (-1.6,0)--(0,0);
\draw (-1.6,-3.5)--(0,-3.5);
\draw (0,0)--(0.5,-1.5);
\draw (0,-3.5)--(0.5,-2);

\draw (0.5,-2.25) rectangle (2.5,-1.25);
\node at (1.5,-1.75){AB};

\draw[->] (2.5,-1.75)--(3.5,-1.75);
\draw[fill=yellow!20] (3.5,-1.4) rectangle (4.5,-2.1);
\node at (4,-1.75){\footnotesize PS};

\draw (4.5,-1.7)--(5.5,0);
\draw[->] (5.5,0)--(7,0);
\draw (4.5,-1.8)--(5.5,-3.5);
\draw[->] (5.5,-3.5)--(7,-3.5);
\node at (6.25,0.4){$\sqrt{\psi}$};
\node at (6.25,-3.9){$\sqrt{1-\psi}$};

\node at (7.5,0){ID};
\node at (7.5,-3.5){EH};;
\end{tikzpicture}

%% file: Rate1.tex
\begin{tikzpicture}

\begin{axis}[%
xmin=0,
xmax=14,
xlabel={$\nicefrac{P_\text{B}}{P}$},
xmajorgrids,
xtick={0,2,4,6,8,10,12,14},
ymin=0,
ymax=2,
ylabel={Achievable rate (bits/cu)},
ylabel near ticks,
ymajorgrids,
ytick={0,0.2,0.4,0.6,0.8,1,1.2,1.4,1.6,1.8,2},
legend style={at={(axis cs:14,2)},anchor= north east, draw=black,fill=white, fill opacity=0.8,legend cell align=left}
]
\addplot [color=red,dashed,mark=square,mark options={solid}]
  table[row sep=crcr]{0	1.01664853279111\\
  1	0.660668218863326\\
  2	0.463693156528066\\
  3	0.368197538520657\\
  4	0.306189584022008\\
  5	0.262105777859003\\
  6	0.229141655777778\\
  7	0.203554984619471\\
  8	0.183115424171839\\
  9	0.166410227703622\\
  10	0.152500763311652\\
  11	0.140508885578685\\
  12	0.130752223992867\\
  13	0.122263310152297\\
  14	0.114810005217804\\
  };
\addlegendentry{Worst-case, $\psi_k=0.3,\ \forall k$}

\addplot [color=green!50!black,solid,mark=square,mark options={solid}]
  table[row sep=crcr]{0	1.01664853279112\\
  1	0.943662089017941\\
  2	0.892626861401986\\
  3	0.853152268711947\\
  4	0.822159544956097\\
  5	0.796603259207287\\
  6	0.774717066787993\\
  7	0.754587334012815\\
  8	0.737145169115885\\
  9	0.720961085501068\\
  10	0.70768875531039\\
  11	0.693262672326653\\
  12	0.681118874650048\\
  13	0.669712065612356\\
  14	0.658612335474566\\
  };
\addlegendentry{Average, $\psi_k=0.3,\ \forall k$}

\addplot [color=red,dashed,mark=asterisk,mark options={solid}]
  table[row sep=crcr]{0	1.50908772555058\\
  1	1.06202964772342\\
  2	0.592321959834248\\
  3	0.411688548174178\\
  4	0.329631121632084\\
  5	0.283348184868092\\
  6	0.243993671760758\\
  7	0.215970445171711\\
  8	0.193729814872342\\
  9	0.174711059337605\\
  10	0.15944160502325\\
  11	0.146628797860188\\
  12	0.137234633460953\\
  13	0.127912589088805\\
  14	0.119777122052303\\
  };
\addlegendentry{Worst-case, $\psi_k=0.6,\ \forall k$}

\addplot [color=green!50!black,solid,mark=asterisk,mark options={solid}]
  table[row sep=crcr]{0	1.50908772555025\\
  1	1.35141733478777\\
  2	1.2560723534339\\
  3	1.18479702754058\\
  4	1.13048810186396\\
  5	1.08607026175188\\
  6	1.04700843398008\\
  7	1.01126579389034\\
  8	0.983132724806032\\
  9	0.957458688771373\\
  10	0.934101317054294\\
  11	0.912604146953446\\
  12	0.891839388040297\\
  13	0.871510469823602\\
  14	0.855766678876369\\
  };
\addlegendentry{Average, $\psi_k=0.6,\ \forall k$}

\addplot [color=red,dashed,mark=o,mark options={solid}]
  table[row sep=crcr]{0	1.81609620162764\\
  1	1.23919554534703\\
  2	0.711632836900461\\
  3	0.428579756622331\\
  4	0.343310703223896\\
  5	0.290224431973288\\
  6	0.251067100476723\\
  7	0.221744353235415\\
  8	0.198361208733517\\
  9	0.179444930658235\\
  10	0.163825900185292\\
  11	0.148703297840223\\
  12	0.139540787212364\\
  13	0.129913589651325\\
  14	0.121529791968665\\
  };
\addlegendentry{Worst-case, $\psi_k=0.9,\ \forall k$}

\addplot [color=green!50!black,solid,mark=o,mark options={solid}]
  table[row sep=crcr]{0	1.81609620162768\\
  1	1.59091068292564\\
  2	1.46049024119545\\
  3	1.37000701913069\\
  4	1.29832207302763\\
  5	1.24175071481241\\
  6	1.18930448719968\\
  7	1.14936149166099\\
  8	1.11259438906469\\
  9	1.07731934701705\\
  10	1.04883459938883\\
  11	1.02344051955916\\
  12	0.997884924635015\\
  13	0.974804641116386\\
  14	0.951521056999655\\
  };
\addlegendentry{Average, $\psi_k=0.9,\ \forall k$}

\end{axis}
\end{tikzpicture}%

%% file: Rate2.tex
\begin{tikzpicture}

\begin{axis}[%
xmin=0,
xmax=14,
xlabel={$\nicefrac{P_\text{B}}{P}$},
xmajorgrids,
xtick={0,2,4,6,8,10,12,14},
ymin=0.4,
ymax=2,
ylabel={Achievable rate (bits/cu)},
ylabel near ticks,
ymajorgrids,
ytick={0.4,0.6,0.8,1,1.2,1.4,1.6,1.8,2},
legend style={at={(axis cs:14,2)},anchor= north east, draw=black,fill=white, fill opacity=0.8, legend cell align=left}
]
\addplot [color=green!50!black,solid,mark=square,mark options={solid}]
  table[row sep=crcr]{0	1.01664853279112\\
  1	0.943662089017941\\
  2	0.892626861401986\\
  3	0.853152268711947\\
  4	0.822159544956097\\
  5	0.796603259207287\\
  6	0.774717066787993\\
  7	0.754587334012815\\
  8	0.737145169115885\\
  9	0.720961085501068\\
  10	0.70768875531039\\
  11	0.693262672326653\\
  12	0.681118874650048\\
  13	0.669712065612356\\
  14	0.658612335474566\\
  };
\addlegendentry{Rec. Struc. 1, $\psi_k=0.3,\ \forall k$}

\addplot [color=red,dashed,mark=square,mark options={solid}]
  table[row sep=crcr]{0	0.952046776777154\\
  1	0.884330564840251\\
  2	0.827095795203666\\
  3	0.777113850711505\\
  4	0.735465957806268\\
  5	0.695601101389132\\
  6	0.661714708048335\\
  7	0.632479115322222\\
  8	0.603499256706302\\
  9	0.578338996484566\\
  10	0.557098388029695\\
  11	0.536194878560808\\
  12	0.516041449762755\\
  13	0.497186250294711\\
  14	0.480758531458103\\
  };
\addlegendentry{Rec. Struc. 2, $\psi_k=0.3,\ \forall k$}

\addplot [color=green!50!black,solid,mark=asterisk,mark options={solid}]
  table[row sep=crcr]{0	1.50908772555025\\
  1	1.35141733478777\\
  2	1.2560723534339\\
  3	1.18479702754058\\
  4	1.13048810186396\\
  5	1.08607026175188\\
  6	1.04700843398008\\
  7	1.01126579389034\\
  8	0.983132724806032\\
  9	0.957458688771373\\
  10	0.934101317054294\\
  11	0.912604146953446\\
  12	0.891839388040297\\
  13	0.871510469823602\\
  14	0.855766678876369\\
  };
\addlegendentry{Rec. Struc. 1, $\psi_k=0.6,\ \forall k$}

\addplot [color=red,dashed,mark=asterisk,mark options={solid}]
  table[row sep=crcr]{0	1.33270793364083\\
  1	1.20055870430343\\
  2	1.09507348294448\\
  3	1.0109500033057\\
  4	0.939157039808953\\
  5	0.877768521404836\\
  6	0.82216949343415\\
  7	0.776148833428963\\
  8	0.736896853082805\\
  9	0.699261561801103\\
  10	0.668863144154364\\
  11	0.635970450578023\\
  12	0.608538169105025\\
  13	0.585609015376016\\
  14	0.562560722371954\\
  };
\addlegendentry{Rec. Struc. 2, $\psi_k=0.6,\ \forall k$}

\addplot [color=green!50!black,solid,mark=o,mark options={solid}]
  table[row sep=crcr]{0	1.81609620162768\\
  1	1.59091068292564\\
  2	1.46049024119545\\
  3	1.37000701913069\\
  4	1.29832207302763\\
  5	1.24175071481241\\
  6	1.18930448719968\\
  7	1.14936149166099\\
  8	1.11259438906469\\
  9	1.07731934701705\\
  10	1.04883459938883\\
  11	1.02344051955916\\
  12	0.997884924635015\\
  13	0.974804641116386\\
  14	0.951521056999655\\
  };
\addlegendentry{Rec. Struc. 1, $\psi_k=0.9,\ \forall k$}

\addplot [color=red,dashed,mark=o,mark options={solid}]
  table[row sep=crcr]{0	1.54518804183097\\
  1	1.36884952393651\\
  2	1.23265705003085\\
  3	1.12524683462025\\
  4	1.03579525258813\\
  5	0.965986412491662\\
  6	0.902746741506528\\
  7	0.845701122472279\\
  8	0.796074440244774\\
  9	0.755634470893558\\
  10	0.714371286314527\\
  11	0.680124308592658\\
  12	0.649985648319117\\
  13	0.622565654917604\\
  14	0.597777474684777\\
  };
\addlegendentry{Rec. Struc. 2, $\psi_k=0.9,\ \forall k$}

\end{axis}
\end{tikzpicture}%

%% file: Energy1.tex
\begin{tikzpicture}

\begin{axis}[%
xmin=0,
xmax=14,
xlabel={$\nicefrac{P_\text{B}}{P}$},
xmajorgrids,
xtick={0,2,4,6,8,10,12,14},
ymin=0,
ymax=10,
ylabel={Harvested energy (dB)},
ylabel near ticks,
ymajorgrids,
ytick={0,2,4,6,8,10},
legend style={at={(axis cs:14,0)},anchor= south east, draw=black,fill=white, fill opacity=0.8, legend cell align=left}
]

\addplot [color=green!50!black,solid,mark=square,mark options={solid}]
  table[row sep=crcr]{0	4.01590886055144\\
  1	4.52291827595852\\
  2	4.87037780422006\\
  3	5.16606705096762\\
  4	5.59555857663366\\
  5	6.11419495912365\\
  6	6.68201759848207\\
  7	7.20382303707603\\
  8	7.65956344734931\\
  9	8.09642526641612\\
  10	8.50302953813649\\
  11	8.85266790072983\\
  12	9.18269864377944\\
  13	9.5082113387408\\
  14	9.80120965807228\\
  };
\addlegendentry{Rec. Struc. 1, $\psi_k=0.3,\ \forall k$}

\addplot [color=red,dashed,mark=square,mark options={solid}]
  table[row sep=crcr]{0	5.48389418132934\\
  1	5.8624572022805\\
  2	6.21003427095872\\
  3	6.53142837817796\\
  4	6.83340598087656\\
  5	7.11541267983445\\
  6	7.381859342626\\
  7	7.62935134892833\\
  8	7.86156572858735\\
  9	8.09072623220261\\
  10	8.29985086916496\\
  11	8.50659029477668\\
  12	8.69023077058881\\
  13	8.88238420805394\\
  14	9.06140952447705\\
  };
\addlegendentry{Rec. Struc. 2, $\psi_k=0.3,\ \forall k$}

\addplot [color=green!50!black,solid,mark=asterisk,mark options={solid}]
  table[row sep=crcr]{0	1.02752122078572\\
  1	1.69402381823995\\
  2	2.08112193545349\\
  3	2.44692023878053\\
  4	2.99479619418398\\
  5	3.59873596563327\\
  6	4.17957654321832\\
  7	4.74266892401469\\
  8	5.19620020591998\\
  9	5.64343753510081\\
  10	6.04569816046571\\
  11	6.41551325997658\\
  12	6.74286149559684\\
  13	7.06495050356767\\
  14	7.34467513889975\\
  };
\addlegendentry{Rec. Struc. 1, $\psi_k=0.6,\ \forall k$}

\addplot [color=red,dashed,mark=asterisk,mark options={solid}]
  table[row sep=crcr]{0	3.05351369446698\\
  1	3.43382202709658\\
  2	3.77970236889134\\
  3	4.10165238467286\\
  4	4.41487038530474\\
  5	4.67274076948809\\
  6	4.95121837019557\\
  7	5.17911134730611\\
  8	5.42657392051862\\
  9	5.64281541568373\\
  10	5.87146148496487\\
  11	6.05794473404498\\
  12	6.26433033376295\\
  13	6.43704788978555\\
  14	6.63677901442064\\
  };
\addlegendentry{Rec. Struc. 2, $\psi_k=0.6,\ \forall k$}

\end{axis}
\end{tikzpicture}%

%% file: Energy2.tex
\begin{tikzpicture}

\begin{axis}[%
xmin=0,
xmax=14,
xlabel={$\nicefrac{P_\text{B}}{P}$},
xmajorgrids,
xtick={0,2,4,6,8,10,12,14},
ymin=0,
ymax=16,
ylabel={Harvested energy (dB)},
ylabel near ticks,
ymajorgrids,
ytick={0,2,4,6,8,10,12,14,16},
legend style={at={(axis cs:14,0)},anchor= south east, draw=black,fill=white, fill opacity=0.8, legend cell align=left}
]

\addplot [color=green!50!black,solid,mark=square,mark options={solid}]
  table[row sep=crcr]{0	4.14630672966892\\
  1	6.15621099917956\\
  2	8.03927930773543\\
  3	9.34847976511308\\
  4	10.3530762901561\\
  5	11.1683928515442\\
  6	11.8545718505906\\
  7	12.4469825839477\\
  8	12.9681984575023\\
  9	13.4335103436627\\
  10	13.8537572159143\\
  11	14.2369024921835\\
  12	14.5889690726535\\
  13	14.9146227317765\\
  14	15.217551517221\\
  };
\addlegendentry{Rec. Struc. 1, $\psi_k=0.3,\ \forall k$}

\addplot [color=red,dashed,mark=square,mark options={solid}]
  table[row sep=crcr]{0	5.48389418132934\\
  1	5.86305924334752\\
  2	6.21039189205948\\
  3	6.52846526394571\\
  4	6.83828722615864\\
  5	7.11965110031225\\
  6	7.37703504977374\\
  7	7.62226978693689\\
  8	7.85172973277093\\
  9	8.09197299808419\\
  10	8.31253587130135\\
  11	8.48967957253314\\
  12	8.69671786115017\\
  13	8.88395953649641\\
  14	9.05118516116692\\
  };
\addlegendentry{Rec. Struc. 2, $\psi_k=0.3,\ \forall k$}

\addplot [color=green!50!black,solid,mark=asterisk,mark options={solid}]
  table[row sep=crcr]{0	1.14807677113851\\
  1	3.58716909145516\\
  2	5.51952814101249\\
  3	6.85216569889276\\
  4	7.870460558193\\
  5	8.69476262191759\\
  6	9.38728946396471\\
  7	9.9844231509457\\
  8	10.5092901967568\\
  9	10.9775091271548\\
  10	11.4001253642877\\
  11	11.7852388705888\\
  12	12.1389664647314\\
  13	12.4660406430523\\
  14	12.7701981583246\\
  };
\addlegendentry{Rec. Struc. 1, $\psi_k=0.6,\ \forall k$}

\addplot [color=red,dashed,mark=asterisk,mark options={solid}]
  table[row sep=crcr]{0	3.05351369446698\\
  1	3.43353838460779\\
  2	3.78088398381578\\
  3	4.1032590579695\\
  4	4.40418472212718\\
  5	4.67822613277088\\
  6	4.94445188263484\\
  7	5.17647699836307\\
  8	5.44155920463447\\
  9	5.65586219521768\\
  10	5.85137909268615\\
  11	6.06623689853883\\
  12	6.27459444836898\\
  13	6.44954733662424\\
  14	6.62087408044807\\
  };
\addlegendentry{Rec. Struc. 2, $\psi_k=0.6,\ \forall k$}

\end{axis}
\end{tikzpicture}%

%% file: IdEhMimo_AkAs.bbl
\begin{thebibliography}{10}
\providecommand{\url}[1]{#1}
\csname url@samestyle\endcsname
\providecommand{\newblock}{\relax}
\providecommand{\bibinfo}[2]{#2}
\providecommand{\BIBentrySTDinterwordspacing}{\spaceskip=0pt\relax}
\providecommand{\BIBentryALTinterwordstretchfactor}{4}
\providecommand{\BIBentryALTinterwordspacing}{\spaceskip=\fontdimen2\font plus
\BIBentryALTinterwordstretchfactor\fontdimen3\font minus
  \fontdimen4\font\relax}
\providecommand{\BIBforeignlanguage}[2]{{%
\expandafter\ifx\csname l@#1\endcsname\relax
\typeout{** WARNING: IEEEtran.bst: No hyphenation pattern has been}%
\typeout{** loaded for the language `#1'. Using the pattern for}%
\typeout{** the default language instead.}%
\else
\language=\csname l@#1\endcsname
\fi
#2}}
\providecommand{\BIBdecl}{\relax}
\BIBdecl

\bibitem{Baszynski2016}
M.~Baszynski, A.~Ruszczyk, and P.~Rydygier, ``{Wireless energy transfer for
  industrial applications: Theory, available solutions and perspectives},'' in
  \emph{2016 International Conference on Signals and Electronic Systems
  (ICSES)}, Sept 2016, pp. 195--199.

\bibitem{Krikidis2014}
I.~Krikidis, S.~Timotheou, S.~Nikolaou, G.~Zheng, D.~W.~K. Ng, and R.~Schober,
  ``{Simultaneous Wireless Information and Power Transfer in Modern
  Communication Systems},'' \emph{invited, IEEE Communications Magazine, Green
  Communications and Computing Networks Series}, vol.~52, no.~11, pp. 104--110,
  Nov. 2014.

\bibitem{Shi2014}
Q.~Shi, L.~Liu, W.~Xu, and R.~Zhang, ``{Joint Transmit Beamforming and Receive
  Power Splitting for {MISO SWIPT} Systems},'' \emph{IEEE Transactions on
  Wireless Communications}, vol.~13, no.~6, pp. 3269--3280, June 2014.

\bibitem{Hameed2014}
Z.~Hameed and K.~Moez, ``{Hybrid Forward and Backward Threshold-Compensated
  RF-DC Power Converter for RF Energy Harvesting},'' \emph{IEEE Journal on
  Emerging and Selected Topics in Circuits and Systems}, vol.~4, no.~3, pp.
  335--343, Sept 2014.

\bibitem{Telatar1999}
E.~Telatar, ``{Capacity of multi-antenna Gaussian channels},'' \emph{European
  Transactions on Telecommunications, ETT}, vol.~10, no.~6, Nov. 1999.

\bibitem{Zhang2013}
R.~Zhang and C.~K. Ho, ``{MIMO Broadcasting for Simultaneous Wireless
  Information and Power Transfer},'' \emph{IEEE Transactions on Wireless
  Communications}, vol.~12, no.~5, pp. 1989--2001, May 2013.

\bibitem{Liu2013}
L.~Liu, R.~Zhang, and K.~C. Chua, ``{Wireless Information and Power Transfer: A
  Dynamic Power Splitting Approach},'' \emph{IEEE Transactions on
  Communications}, vol.~61, no.~9, pp. 3990--4001, September 2013.

\bibitem{Kariminezhad2017SPL}
A.~Kariminezhad, S.~Gherekhloo, and A.~Sezgin, ``{Optimal Power Splitting for
  Simultaneous Information Detection and Energy Harvesting},'' \emph{IEEE
  Signal Processing Letters}, vol.~24, no.~7, pp. 963--967, July 2017.

\bibitem{Ye2003}
S.~Ye and R.~S. Blum, ``{Optimized signaling for MIMO interference systems with
  feedback},'' \emph{IEEE Transactions on Signal Processing}, vol.~51, no.~11,
  pp. 2839--2848, Nov 2003.

\bibitem{Jorswieck2004}
E.~Jorswieck and H.~Boche, ``{Performance Analysis of Capacity of MIMO Systems
  under Multiuser Interference Based on Worst-Case Noise Behavior},''
  \emph{EURASIP Journal on Wireless Communications and Networking}, vol. 2004,
  no.~2, p. 670321, Dec 2004.

\bibitem{Fan1953}
K.~Fan, ``{Minimax theorems},'' \emph{Proceedings of the National Academy of
  Sciences of the United States of America}, vol.~39, pp. 42--47, 1953.

\end{thebibliography}
